# Spin-charge coupling and decoupling in perovskite-type iron oxides (Sr$_{1-x}$Ba$_x$)$_{2/3}$La$_{1/3}$FeO$_3$

M. Onose[1,2], H. Takahashi[2,3], T. Saito[4,5], T. Kamiyama[4], R. Takahashi[6], H. Wadati[6,7], S. Kitao[8], M. Seto[8], H. Sagayama[9], Y. Yamasaki[10], T. Sato[11,12,13], F. Kagawa[1,11] and S. Ishiwata[2,3]

[1] *Department of Applied Physics, University of Tokyo, Bunkyo-ku, Tokyo 113-8656, Japan*
[2] *Division of Materials Physics and Center for Spintronics Research Network (CSRN), Graduate School of Engineering Science, Osaka University, Toyonaka, Osaka 560-8531, Japan*
[3] *Division of Spintronics Research Network, Institute for Open and Transdisciplinary Research Initiatives, Osaka University, Yamadaoka 2-1, Suita, Osaka 565-0871, Japan*
[4] *Institute of Materials Structure Science (IMSS), High Energy Accelerator Research Organization (KEK), Tokai, Ibaraki 319-1106, Japan*
[5] *The Graduate University for Advanced Studies, SOKENDAI, Tokai, Ibaraki 319-1106, Japan*
[6] *Department of Material Science, Graduate School of Science, University of Hyogo, Ako, Hyogo 678-1297, Japan*
[7] *Institute of Laser Engineering, Osaka University, Suita, Osaka 565-0871, Japan*
[8] *Institute for Integrated Radiation and Nuclear Science, Kyoto University, Kumatori, Osaka 590-0494, Japan*
[9] *Institute of Materials Structure Science (IMSS), High Energy Accelerator Research Organization (KEK), Tsukuba, Ibaraki 305-0801, Japan*
[10] *National Institute for Materials Science (NIMS), Tsukuba, Ibaraki 305-0047, Japan*
[11] *RIKEN Center for Emergent Matter Science (CEMS), Wako, Saitama 351-0198, Japan*
[12] *Research Center of Integrative Molecular Systems (CIMoS), Institute for Molecular Science, Myodaiji, Okazaki 444-8585, Japan*
[13] *The Graduate University for Advanced Studies, SOKENDAI, Myodaiji, Okazaki 444-8585, Japan*

**Abstract**

The perovskite-type iron oxide Sr$_{2/3}$La$_{1/3}$FeO$_3$ is known to show characteristic spin-charge ordering (SCO), where sixfold collinear spin ordering and threefold charge ordering are coupled with each other. Here, we report the discovery of a spin-charge decoupling and an antiferromagnetic (AFM) state competing with the SCO phase in perovskites (Sr$_{1-x}$Ba$_x$)$_{2/3}$La$_{1/3}$FeO$_3$. By comprehensive measurements including neutron diffraction, Mössbauer spectroscopy, and x-ray absorption spectroscopy, we found that the isovalent Ba$^{2+}$ substitution systematically reduces the critical temperature of the SCO phase and additionally yields the spin-charge decoupling in $x > 0.75$. Whereas the ground state remains in the SCO phase in the whole $x$ region, an unexpected G-type AFM phase with incoherent charge ordering or charge fluctuation appears as the high-temperature phase in the range of $x > 0.75$. Reflecting the competing nature between them, the G-type AFM phase partially exists as a metastable state in the SCO phase at low temperatures. We discuss the origin of the spin-charge decoupling and the emergence of the G-type AFM phase with charge fluctuation in terms of the bandwidth reduction by the Ba substitution.

## I. INTRODUCTION

Spin and charge ordering in transition-metal oxides manifests as an emergent collective phenomenon of strongly correlated electrons [1,2]. The coupling between spin and charge degrees of freedom plays a key role in the emergence of a rich variety of quantum phases with exotic properties [3]. This is exemplified by the stripe phase in layered cuprates showing high-temperature superconductivity [4,5], the CE-type phase in perovskite-type manganites showing the colossal magnetoresistance effect [6,7], and the Verwey transition in a magnetite showing a large magnetoelectric effect [8,9].

In perovskite-type iron oxides $A$FeO$_3$ containing unusually-high-valence Fe$^{4+}$ ions, unique types of spin and charge ordering emerge, depending on the A-site cation [10]. For instance, SrFeO$_3$ hosts a rich variety of incommensurate helimagnetic phases [11,12], whereas CaFeO$_3$ shows helimagnetic ordering and charge disproportionation described by 2Fe$^{4+}$ → Fe$^{3+}$ + Fe$^{5+}$, which are decoupled with each other [13,14]. On the other hand, Sr$_{2/3}$La$_{1/3}$FeO$_3$ with Fe$^{3.67+}$ ions shows a characteristic spin-charge ordering (SCO) where threefold charge ordering, 3Fe$^{3.67+}$ → 2Fe$^{3+}$ + Fe$^{5+}$, and sixfold collinear magnetic ordering are coupled with each other, both of which propagate along the [111] direction of the pseudocubic unit cell [15,16]. The sixfold collinear magnetic structure in Sr$_{2/3}$La$_{1/3}$FeO$_3$ can be reasonably explained by spin-charge coupling in the Fe$^{3.67+}$ state, i.e., the antiferromagnetic (AFM) and ferromagnetic (FM) interactions are favored for Fe$^{3+}$ − Fe$^{3+}$ and Fe$^{3+}$ − Fe$^{5+}$, respectively [see Fig. 1(c)] [17,18]. However, recent experimental reports have cast the possibility that magnetic and charge ordering can be decoupled by A-site substitution. For instance, in Sr$_{1-x}$La$_x$FeO$_3$, the SCO phase was found to spread in the wide range around $x$ ~ 1/3, which is apparently connected to the presumably single-$q$ incommensurate helimagnetic phase at $x = 0$ [12,19]. In LaCa$_2$Fe$_3$O$_9$, whereas it hosts the SCO-type ground state, the coexistence of charge ordering propagating along [010] and a helical spin structure with a propagation vector $q$ = (1/3, 1/3, 1/3) on the pseudocubic unit cell was found in the high-temperature phase [20,21].

In this paper, we report the emergence of a G-type AFM phase, where spin and charge are decoupled to yield a charge fluctuating state in (Sr$_{1-x}$Ba$_x$)$_{2/3}$La$_{1/3}$FeO$_3$ with $x > 0.75$. While the SCO phase seems to remain a ground state in all $x$ regions, the G-type AFM phase appears as a high-temperature phase competing with the SCO phase. Based on comprehensive measurements, we discuss the origin of incoherent charge ordering or charge fluctuation in the G-type AFM phase in terms of the incommensurability between the spin and charge ordering.

## II. EXPERIMENTAL METHODS

Polycrystalline samples of (Sr$_{1-x}$Ba$_x$)$_{2/3}$La$_{1/3}$FeO$_3$ ($x$ = 0, 0.25, 0.5, 0.75, 0.9, 1) were synthesized by the following procedure. First, the starting materials (SrCO$_3$, BaCO$_3$, La$_2$O$_3$, and Fe$_2$O$_3$) were stoichiometrically mixed and sintered at 1100 − 1250 ℃ for 2 − 3 times with intermediate grindings. Then the sample was annealed at 1300 ℃ in a flow of Ar for 48 h. Finally, the obtained powder with oxygen-deficient perovskite-type structure was sealed in a gold capsule with an oxidizing agent (NaClO$_3$), which was annealed at 500 ℃ and 8 GPa for 1 h by using a cubic-anvil-type high-pressure apparatus.

Synchrotron powder x-ray diffraction (SXRD) was performed at beamline 8A, Photon Factory, KEK, Japan. The wavelength of synchrotron x-ray (0.68975 − 0.68980 Å) was calibrated for each composition using a CeO$_2$ standard. The crystal structure was refined by Rietveld analysis using RIETAN-FP [22]. Powder neutron diffraction (ND) data for Ba$_{2/3}$La$_{1/3}$FeO$_3$ ($x$ = 1) were collected with a time-of-flight diffractometer at BL09 (SPICA), J-PARC MLF, Japan. The crystal and magnetic structures were analyzed by the Rietveld method using Z-Rietveld [23,24]. The $^{57}$Fe Mössbauer spectra for $x$ = 1 were measured with a $^{57}$Co source in Rh. The velocity was calibrated with $\alpha$-Fe. Fits to the Mössbauer spectra were performed by the least-squares method assuming Lorentzian peaks. The extended x-ray absorption fine structure (EXAFS) spectra at the Fe K-edge were measured in a transmission mode at BL11 of SAGA Light Source in Kyushu Synchrotron Light Research Center.

### III. RESULTS AND DISCUSSION

The SXRD patterns at $T$ = 300 K were well fitted assuming a rhombohedral perovskite-type structure ($R\bar{3}c$) for $x$ = 0 and a simple cubic perovskite-type structure ($Pm\bar{3}m$) for 0.25 ≤ $x$ ≤ 1, respectively. The results of Rietveld analyses are shown in Fig. S1 and Table S1 in the Supplemental Material [25]. The temperature dependence of the SXRD patterns for $x$ = 0.25 indicates that it undergoes a structural transition from the high-temperature cubic to the low-temperature rhombohedral phase between $T$ = 270 and 250 K (see Fig. S2 in the Supplemental Material [25]). As shown in Fig. 1(a), the pseudocubic lattice constant $a_{\text{cubic}}$ for (Sr$_{1-x}$Ba$_x$)$_{2/3}$La$_{1/3}$FeO$_3$ at $T$ = 300 K changes linearly with $x$ following Vegard's law, indicating the complete formation of a solid solution with negligible oxygen deficiency.

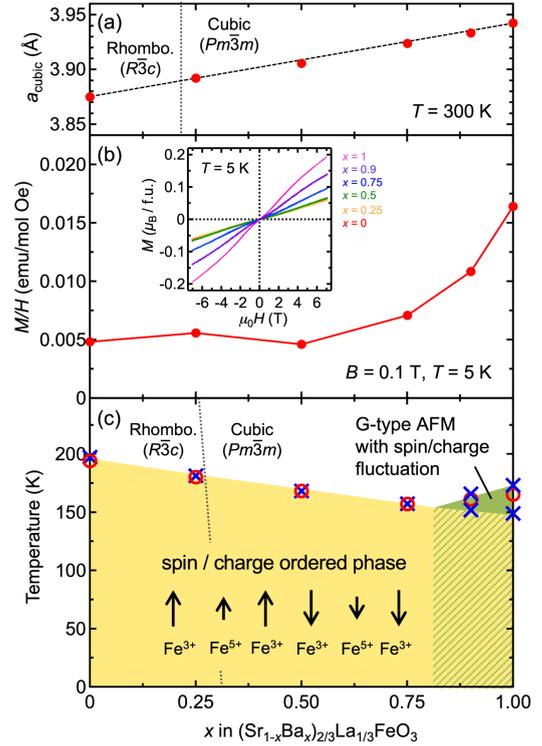

Fig. 1. (a) The pseudocubic lattice constant $a_{\text{cubic}}$, (b) the magnetic susceptibility at $T$ = 5 K, and (c) the phase diagram of (Sr$_{1-x}$Ba$_x$)$_{2/3}$La$_{1/3}$FeO$_3$ plotted as a function of $x$. $a_{\text{cubic}}$ for $x$ = 0 was calculated by $a_{\text{cubic}}$ = $(V/6)^{1/3}$, where $V$ represents the volume of the rhombohedral unit cell. The inset in (b) shows the magnetization curves at 5 K. The red open circles and blue crosses in (c) represent the transition temperatures determined by the resistivity and magnetization measurements, respectively. The broken line in (c) represents the structural phase boundary between the rhombohedral and cubic phases.

Figures 2(a) and 2(b) show the temperature dependence of resistivity and magnetization, respectively. The resistivity for 0 ≤ $x$ ≤ 0.75 shows a first-order metal-insulator transition associated with the charge ordering. This transition is detectable also as the anomaly in the magnetization, indicating the emergence of the SCO phase in the wide $x$ region, as shown in Fig. 1(c). Notably, the magnetization for 0.9 ≤ $x$ ≤ 1 shows two-step transitions at $T_{N1}$ and $T_{N2}$ (= 149 and 173 K at $x$ = 1, respectively), of which the lower-temperature one at $T_{N1}$ is first order. Given that the first-order transition at $T_{N1}$ reflects the onset of the SCO phase, this result indicates the emergence of a new phase in between $T_{N1}$ and $T_{N2}$, which will be verified later from the data of powder ND and Mössbauer spectroscopy. Although the resistivity for 0.9 ≤ $x$ ≤ 1 shows semiconducting behavior in the entire temperature range, a slight kink corresponding to the emergence of the new phase is found around $T_{N2}$, as shown in Fig. 2(c), which can be associated with the development of incoherent charge ordering or charge fluctuation, as discussed later. The onset of the new phase at $T_{N2}$ is also apparent in the specific heat data, as shown in Fig. 2(e).

To get insight into the phase transition at $T_{N1}$, we examined the cooling-rate dependence of magnetization for $x$ = 0.9, as shown in Fig. 2(d). A cooling-rate-dependent bifurcation is observed below $T_{N1}$, suggesting the coexistence of the SCO phase and the new phase. Reflecting the strong first-order nature of this transition, the new high-temperature phase tends to partially exist as a metastable state upon cooling, whose volume fraction seems to increase with a faster cooling rate. The enhancement of magnetic susceptibility at 5 K and the nonlinearity of the magnetization curves for $x \geq 0.9$ [see Fig. 1(b)] can be also associated with the phase coexistence, as it allows a multidomain state with spin fluctuation near the domain boundary.

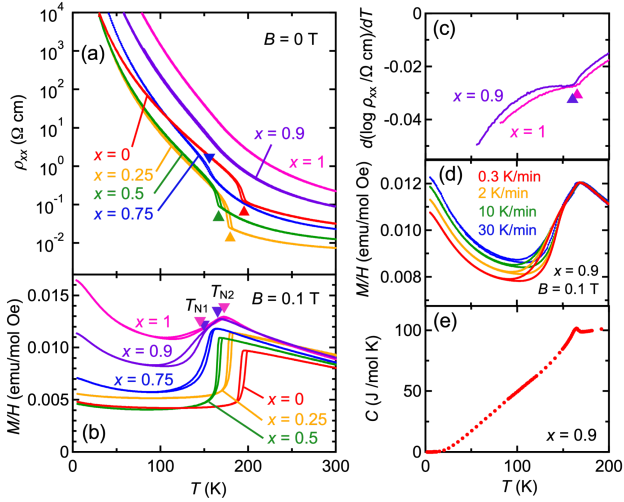

Fig. 2. Temperature dependence of (a) resistivity and (b) magnetic susceptibility for $(Sr_{1-x}Ba_x)_{2/3}La_{1/3}FeO_3$. (c) Temperature derivative of the logarithm of resistivity for $x = 0.9$ and 1. (d) Magnetic susceptibility measured at various cooling rates for $x = 0.9$. (e) Specific heat as a function of temperature for $x = 0.9$.

Figures 3(a) and 3(b) show the powder ND patterns for $Ba_{2/3}La_{1/3}FeO_3$ ($x = 1$) at selected temperatures. The ND patterns for the new phase (150 K ≤ $T$ ≤ 170 K) and the low-temperature phase (4 K ≤ $T$ ≤ 140 K) can be well indexed with the G-type AFM structure and the six-fold collinear magnetic structure corresponding to the SCO phase, respectively. The magnetic structures for both phases are illustrated in Fig. 3(c). The ND patterns for the paramagnetic phase (180 K ≤ $T$ ≤ 300 K) are well fitted with the simple cubic perovskite-type structure ($Pm\bar{3}m$), which is consistent with the SXRD data (see also Table S2 and Fig. S3 in the Supplemental Material [25] for the detailed results of Rietveld refinements of the ND patterns. In Figs. S3(a) and S3(b) in the Supplemental Material [25], the ND patterns shown in Figs. 3(a) and 3(b) are replotted in logarithmic scales, so that weak magnetic reflections appear more clearly.)

Since the coexistence of the SCO phase and the G-type AFM phase is implied below $T_{N1}$ in the magnetization measurements with different cooling rates, we performed Rietveld refinement, assuming the coexistence of these two phases, as shown in Fig. S4 in the Supplemental Material [25]. However, it was technically difficult to estimate the volume fraction of each phase with high accuracy, as the reliability factor $R_{wp}$ shows no improvement by changing the volume fraction. Therefore, we adopted the result of the simplest Rietveld refinement, assuming only the SCO phase, which can be regarded as a major phase, as described below. Figure 3(f) shows the temperature dependence of refined magnetic moments. In the SCO phase below $T_{N1}$, a significant difference is found between the magnetic moments of the two Fe sites, as 2.46 and 0.89 $\mu_B$ at $T = 4$ K, which can be assigned to $Fe^{3+}$ and $Fe^{5+}$, respectively. The refined magnetic moments are apparently smaller than the reported values for $Sr_{2/3}La_{1/3}FeO_3$ ($Fe^{3+}$: 3.67 $\mu_B$, $Fe^{5+}$: 3.26 $\mu_B$) [26], which implies the oversimplification of the analysis, assuming only the SCO phase. On the other hand, as shown in Fig. S4 in the Supplemental Material [25], the refined magnetic moments of the SCO phase become unreasonably large when the assumed volume fraction of the SCO phase becomes as small as 30 %. Therefore, it is reasonable to presume that the G-type AFM phase exists as a minor phase below $T_{N1}$.

Figures 3(d) and 3(e) show the temperature dependence of the intensity and full width at half maximum (FWHM) for the two representative magnetic reflections (#1 and #2), obtained by the profile fitting of each peak (see also Fig. S5 in the Supplemental Material [25]). The magnetic reflection #1 at $d = 3.3$ Å is indexed as the overlapped peaks of the scattering vectors of $(1+\delta, +\delta, +\delta)$ and $(1-\delta, 1-\delta, -\delta)$ ($\delta = 1/6$) as the SCO phase. This reflection cannot be indexed as the G-type AFM phase. On the other hand, the magnetic reflection #2 at $d = 4.55$ Å can be indexed by both phases: a single peak of $(-\delta, -\delta, 1-\delta)$ as the SCO phase and/or overlapped peaks of (1/2, 1/2, 1/2), (1/2, 1/2, -1/2), (1/2, -1/2, 1/2), and (1/2, -1/2, -1/2) as the G-type AFM phase. Considering the temperature dependence of the intensity and FWHM for #1 and #2, #1 below $T_{N1}$ and #2 above $T_{N1}$ can be associated with the SCO phase and the G-type AFM phase, respectively. However, the intensity of #1 appears from slightly above $T_{N1}$. This indicates that there exists significant development of the short-range ordering of SCO even above $T_{N1}$, reflecting the competing nature between the SCO phase and the G-type AFM phase. The competition between these two phases can also be confirmed as the slight increase of FWHMs for magnetic peaks #1 and #2 upon cooling < 100 K, much lower than $T_{N1}$.

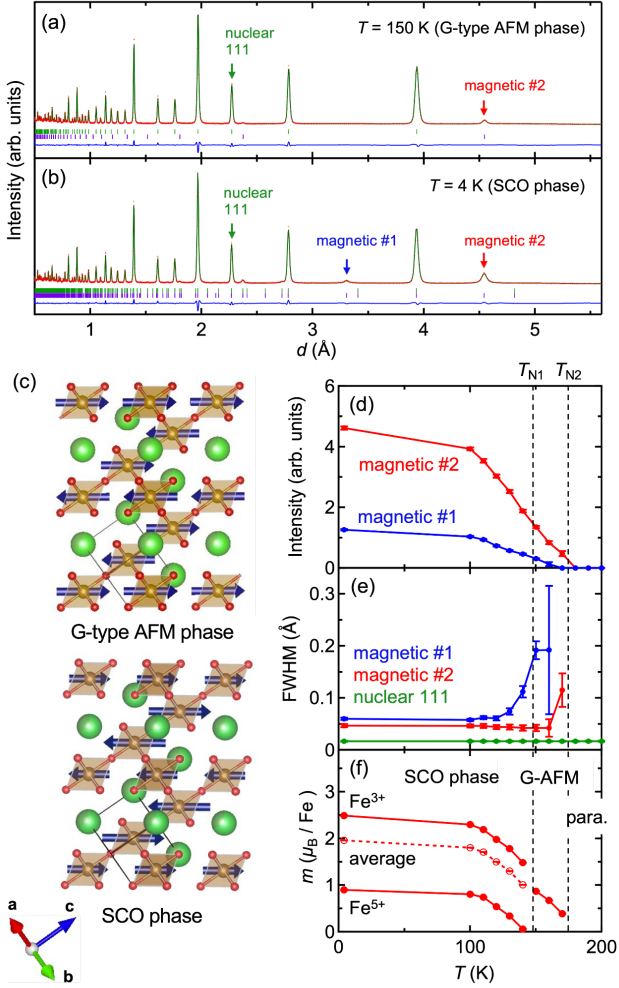

Fig. 3. Powder neutron diffraction patterns and results of Rietveld refinements for $Ba_{2/3}La_{1/3}FeO_3$ ($x = 1$) at (a) 150 K and (b) 4 K. The green and purple ticks indicate the positions of the nuclear and magnetic diffraction peaks, respectively. (c) Schematic images of magnetic structure for the G-type antiferromagnetic (AFM) phase and spin-charge ordering (SCO) phase, visualized by VESTA software [27]. Temperature dependence of the (d) intensity and (e) full width at half maximum (FWHM) of nuclear 111 reflection and two selected magnetic scattering reflections. (The scattering vector of each magnetic reflection is described in the main text.) (f) Temperature dependence of refined magnetic moments.

To further clarify the development of charge ordering in compounds with $0.9 \leq x \leq 1$, Mössbauer spectroscopy measurements were performed for $x = 1$. As shown in Fig. 4(a), the spectra in the paramagnetic phase (180 K $\leq T \leq$ 300 K) can be well fitted by quadrupole-splitted doublets for $Fe^{3.67+}$, and those below $T_{N1}$ (77 K $\leq T \leq$ 140 K) by the superposition of two sextets corresponding to $Fe^{3+}$ and $Fe^{5+}$, indicating the emergence of the SCO phase, as suggested by the ND measurements. The area ratio for $Fe^{3+}$ and $Fe^{5+}$ are 62.5 and 37.5 % at 77 K, respectively, which is close to the expected ratio of 2:1. Although the spectra at $T = 160$ K in the G-type AFM phase are significantly broadened, making it difficult to obtain a satisfactory fitting, the asymmetric spectral shape, i.e., an asymmetric distribution of isomer shift (IS), suggests the development of incoherent charge ordering or the slowing down of charge fluctuation as $3Fe^{3.67+} \rightleftarrows$ $2Fe^{3+} + Fe^{5+}$. This charge fluctuation would be as slow as the laboratory timescale around and below $T_{N2}$, considering the observation of the weak anomaly in the temperature derivative of resistivity near $T_{N2}$ [Fig. 2(c)] and the abrupt change in Fe-O bond length slightly above $T_{N2}$ revealed by EXAFS measurements, which is discussed later.

Figures 4(b)-4(d) show the temperature dependence of Mössbauer parameters. The data for $x = 0$ reported in a previous study [28] are also plotted for comparison. The hyperfine fields for $x = 0$ and 1 are comparable with each other, implying that they share the same SCO phase as a ground state. On the other hand, IS for $x = 1$ is systematically larger than that for $x = 0$ by ~0.01 mm/s. Although the increase in IS is typically considered a signature of the decrease in the valence of Fe, the difference in IS between $x = 0$ and 1 is most likely due to the change in the bonding state of Fe-O. The electron density at the nucleus position is expected to be smaller for $x = 1$ than 0 because of the following two reasons: (1) the broadening of the 4s electron distribution with the increase in the lattice constant and (2) the enhancement of screening effect by the localization of 3d electrons. In fact, the ISs of the related $Fe^{4+}$ compounds $SrFeO_3$ and $BaFeO_3$ are reported as 0.07 mm/s [29] and 0.146 mm/s [30], respectively, whose difference is comparable with that between $x = 0$ and 1 in $(Sr_{1-x}Ba_x)_{2/3}La_{1/3}FeO_3$.

The quadrupole splitting (QS) in the SCO phase is close to zero for both $x = 0$ and 1, as shown in Fig. 4(d), reflecting the high octahedral symmetry in the electronic configuration of $Fe^{3+}(t_{2g}^3 e_g^2)$ and $Fe^{5+}(t_{2g}^3)$. In the paramagnetic phase, the QS for $x = 1$ is significantly larger than that for $x = 0$. The enhancement of QS for $x = 1$ above $T_{N2}$ presumably reflects the abovementioned development of incoherent charge ordering or charge fluctuation causing a finite distribution of IS.

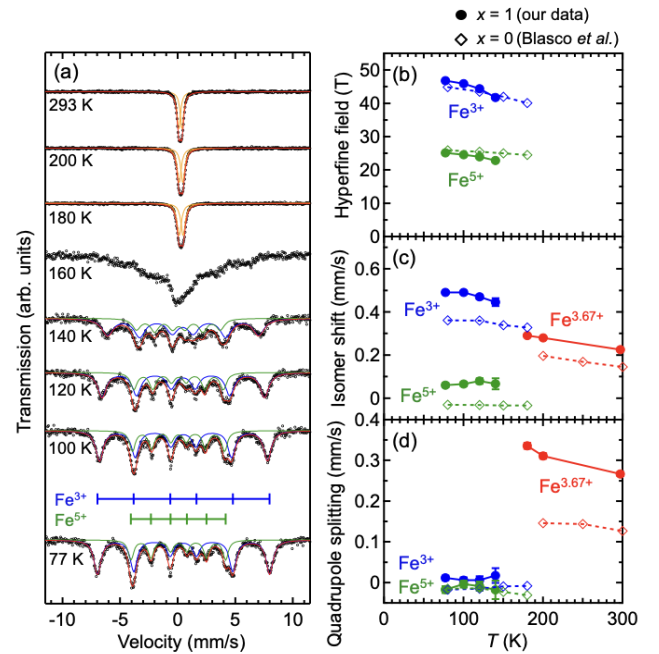

Fig. 4. (a) Temperature dependence of the Mössbauer spectra for $(Sr_{1-x}Ba_x)_{2/3}La_{1/3}FeO_3$ ($x = 1$) with the fitting curves of Lorentzian functions. (b) Hyperfine field, (c) isomer shift, and (d) quadrupole

splitting obtained by the fitting of the Mössbauer spectra. The data for $x = 0$ are plotted for comparison [28].

To characterize the change in the Fe-O bond length associated with the charge ordering, we measured EXAFS (see Fig. S6 in the Supplemental Material [25]), of which Fourier transforms ($|\chi(R)|$) are shown in Figs. 5(a) and 5(b). The peaky structure corresponding to the Fe-O bond can be found ~2 Å. The peak for $x = 1$ is broader than that for $x = 0$ in a wide temperature region [see the inset of Fig. 5(c)], reflecting the large local disorder around the A site accommodating large $Ba^{2+}$ ions and small $La^{3+}$ ions. As shown in Fig. 5(c), the peak position of $|\chi(R)|$ for $x = 0$ shows a significant change ~180 K, slightly below the transition temperature of the SCO phase, as previously reported [31]. On the other hand, for $x = 1$, the peak position of $|\chi(R)|$ exhibits a similar change at temperature higher than $T_{N2}$, which is also a manifestation of the possible development of incoherent charge ordering or the slowing down of charge fluctuation even in the paramagnetic phase as well as in the G-type AFM phase.

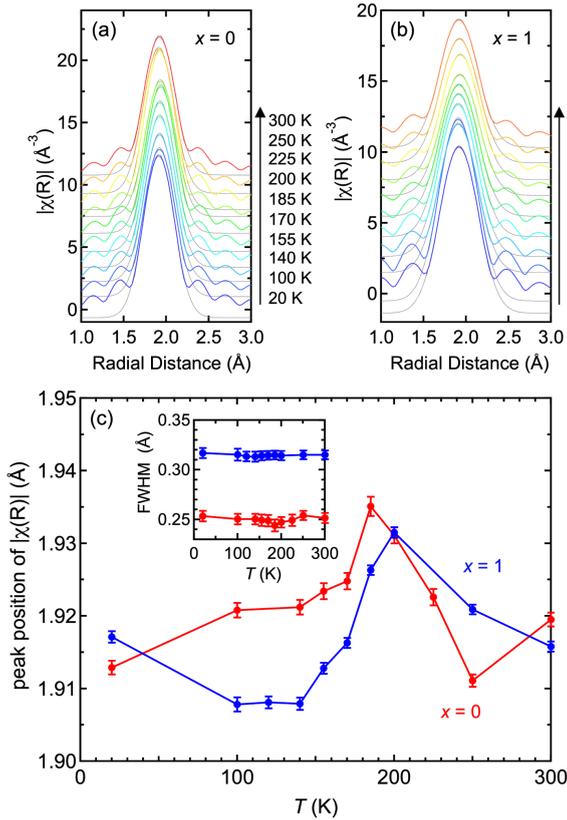

Fig. 5. Fourier transform of the extended x-ray absorption fine structure (EXAFS) spectra for (a) $Sr_{2/3}La_{1/3}FeO_3$ ($x = 0$) and (b) $Ba_{2/3}La_{1/3}FeO_3$ ($x = 1$). The gray solid lines show the Gaussian fitting. (c) Temperature dependence of the peak position of $|\chi(R)|$ for $x = 0$ and 1. The inset shows the full width at half maximum (FWHM) of the peak estimated by the Gaussian fitting.

Finally, we discuss the origin of the emergence of the G-type AFM phase and the spin-charge decoupling in $(Sr_{1-x}Ba_x)_{2/3}La_{1/3}FeO_3$. Given that the bandwidth depends strongly on the Fe-O-Fe bond angle and the Fe-O bond length, the large and nonmonotonic change in the resistivity at 300 K as a function of $x$ can be associated with the variation of $pd\sigma$ dependent on the bandwidth [see Fig. 2(a)]. In the previous Hartree-Fock calculations for $Sr_{2/3}La_{1/3}FeO_3$ [32], the G-type AFM phase is predicted to be rather unstable as compared with the SCO phase (by 480 meV/f.u.). Thus, the observation of the G-type AFM phase as well as the SCO phase in $(Sr_{1-x}Ba_x)_{2/3}La_{1/3}FeO_3$ indicates that the change in lattice structure as a function of $x$ causes a significant change in $pd\sigma$. Regarding the spin-charge decoupling, let us focus on the fact that the resistivity of all compounds increases upon decreasing temperature in the paramagnetic region ~300 K. This presumably reflects the polaronic conduction with strong electron-phonon coupling, which is expected to be enhanced by the bandwidth reduction. Assuming that the SCO transition is dominated by the charge ordering, which tends to be stabilized by the electron-phonon coupling and the bandwidth reduction, the critical temperature of the SCO phase should increase upon the increment of $x$ in $x > 0.25$. Since this is not the case, it is likely that the SCO transition is dominated by the magnetic ordering. Whereas the bandwidth reduction is expected to destabilize the magnetic ordering, it should stabilize the charge ordering. This opposite effect of such a change in bandwidth is expected to destabilize the spin-charge coupling state as $x$ increases from 0, eventually leading to the spin-charge decoupling in $x > 0.75$. Note that the G-type AFM ordering with twofold periodicity cannot couple with the three-fold charge ordering as in the SCO phase. The incommensurability between magnetic ordering and charge ordering likely results in the incoherent charge ordering or charge fluctuation in the G-type AFM phase for $x > 0.75$.

## IV. SUMMARY

To summarize, we established the phase diagram of $(Sr_{1-x}Ba_x)_{2/3}La_{1/3}FeO_3$, which is highlighted by the spin-charge decoupling and the emergence of the G-type AFM phase with incoherent charge ordering or charge fluctuation in the temperature range $T_{N1} < T < T_{N2}$ for $x > 0.75$. Considering the decrease in the transition temperature upon the increment of $x$, we propose the importance of magnetic interaction on the stability of the SCO phase. The contrasting bandwidth dependence on the stability of magnetic ordering and charge ordering is a possible origin of the spin-charge decoupling. The incommensurability between the G-type AFM ordering and threefold charge ordering derived from SCO phase possibly results in the incoherent charge ordering or charge fluctuation. At low temperatures for $x > 0.75$, the G-type AFM phase partially exists as a metastable phase coexisting with the SCO phase, which is supported by the significant cooling rate dependence of magnetization [Fig. 2(d)] and the enhanced magnetization at 5 K [Fig. 1(b)]. The present system offers an ideal opportunity to systematically study the spin-charge interplay in perovskite iron oxides with strong p-d hybridization.

## ACKNOWLEDGMENTS

This work was partly supported by JSPS, KAKENHI (Grants No. 19H05824, No. 19K14652, No. 20H01866, No. 21H01030, No. 22H00343, and No. 22J13408), the Murata Science Foundation and Asahi Glass Foundation. The synchrotron powder XRD and the powder ND measurements were performed with the approvals of the Photon Factory Program Advisory Committee (Proposal No. 2018S2-006) and the Neutron Scattering Program Advisory Committee (Proposal 2019S10). The EXAFS measurements were performed at BL11 of SAGA Light Source (Proposal No. 1904022F).